\documentclass[12pt]{article}
\usepackage{graphics}
\usepackage{amssymb}
\usepackage{color}
\usepackage{amsmath}
\usepackage{amsthm}
\usepackage{amsopn}
\usepackage{graphicx}
\usepackage{axodraw}

\def\p{\partial}
\def\ga{\gamma}

\def\th{\theta}

\def\be{\begin{equation}}
\def\ee{\end{equation}}
\def\bea{\begin{eqnarray}}
\def\eea{\end{eqnarray}}

\begin{document}

 \begin{center}

 {\bf{\large On the Higgs mechanism in $U(1)\times U(2)\times U(3)$ gauge group as a base for noncommutative standard model}}

\vskip 4em

 {{\bf M. M. Ettefaghi}\footnote{email: mettefaghi@qom.ac.ir }
 }
 \vskip 1em
 Department of Physics, University of Qom, Qom 371614-6611,
I.R. Iran.

 \end{center}

 \vspace*{1.9cm}

\begin{abstract}
Noncommutative (NC) space-time leads to some strong constraints on the possible choices of gauge groups  and allowed representations of matter and gauge fields. The standard model based on $U(3)\times U(2)\times U(1)$ can be transcribed to NC space-time consistently with these constraints \cite{nNCSM}. In fact, through two new symmetry reduction mechanisms, this gauge group is reduced to the usual standard model one. In this paper, we show that, on the contrary to the usual standard model, the Goldestone boson equivalence theorem is violated due to the incompatibility of the new symmetry reduction mechanisms with the electroweak Higgs mechanism.

\end{abstract}


\newpage

\section{Introduction}
One of the most important questions in particle physics is related to the UV completion of the standard model (SM). For this purpose, we are interested to construct the SM in noncommutative (NC) space-time because NC geometry appears in string theory with a nonzero B-field \cite{sw,aash1,aash2}. In field theory on NC space-time, one encounters new properties such as  UV/IR
mixing \cite{uv/ir}, Lorentz violation \cite{LV} and CP-violation \cite{CP}. In particular, we have new restrictions for the construction of a gauge theory on NC space-time \cite{nogo}.

In NC space-time, the coordinates of space-time are operators and, in the canonical version, obey from the following algebra:
 \be [\hat{x}^\mu,\hat{x}^\nu]=i\th^{\mu\nu},
\ee
where a hat indicates an NC coordinate and $\th^{\mu\nu}$ is a
real, constant and antisymmetric matrix.
 To construct the NC field theory, according to the Weyl-Moyal correspondence, an ordinary function can
  be used instead of the corresponding NC one by replacing the ordinary product with the star product as follows:
\be\label{starproduct}
 f\star
g(x,\theta)=f(x,\theta)\exp(\frac{i}{2}\overleftarrow{\partial}_\mu
\theta^{\mu\nu}\overrightarrow {\partial}_\nu)g(x,\theta). \ee
This correspondence leads to some new restrictions on the NC gauge theory which are gathered into a no-go theorem \cite{nogo}:
\begin{itemize}
\item The only possible gauge groups in the NC space-time is $U(n)$ denoted by
$U_\star(n)$.
\item The fundamental, anti-fundamental and adjoint representations are allowed for the $U_\star(n)$ gauge group. In particular, for arbitrary fixed charge $q$, only the
matter fields with charges $\pm q$ and zero are permissible  in the $U_\star(1)$ gauge group.
\item When we have a gauge group constructed by the
direct product of several simple gauge groups, a
matter field can be charged under at most two of them.
\end{itemize}
Therefore, the collection of these constraints imposes severe challenges in constructing a consistent and realistic noncommutative standard model (NCSM). However, this problem has been treated with two different strategies. First, the usual standard model (SM) gauge group was transcribed to the NC space-time with two modifications; replacing  the ordinary product between the SM fields by a star one and expanding the NC fields in terms of the ordinary related gauge boson and the corresponding fields via Seiberg-Witten maps \cite{sm}. It has been shown that this approach cannot bypass the NC no-go theorem restrictions \cite{bypass}. Second, the gauge group is restricted
to $U_\star(n)$ but the symmetry group of the SM is
achieved by the reduction of $U_\star(3)\times U_\star(2)\times U_\star(1)$ down to
$NCSU(3)\times NCSU(2)\times U(1)$ through appropriate symmetry
breaking mechanisms \cite{nNCSM}. Since this model includes new degrees of freedom in comparison to the SM, hereafter we will call it non-minimal NC SM (nNCSM). The representations of matter fields in this model are chosen in such a way that the NC no-go theorem is respected (in particular the charge quantization is solved). 
However, it has been shown that the unitarity is violated due to the new symmetry reductions \cite{unitary violation}. For instance, in calculating the amplitude of the process $W^+_LW^-_L\rightarrow W^+_LW^-_L$, although the leading orders, ${\cal O}(s^2/M_W^4)$, are eliminated similar to the usual SM, the sub-leading orders, ${\cal O}(s/M_W^2)$, remain. Of course, it is noticeable that this anomaly is removed in the commutative limit.

In this paper, we ignore the NC modifications for simplicity  and study the compatibility of new symmetry reductions with electroweak Higgs mechanism in nNCSM (It is better we use the abbreviation nSM instead of nNCSM when we ignore NC modifications). The coupling of new gauge bosons in nSM to the Higgs doublet causes them not to be independent from each other. Therefore, there is room for studying the electroweak Higgs mechanism in this theory. The Goldestone boson equivalence theorem (GBET), which is a consequence of spontaneously symmetry breaking, is a criterion of consistency of Higgs mechanism. Hence, we study the GBET for this purpose.

 
We know that through the Higgs mechanism, the unphysical particles well known as Goldestone bosons are eliminated and the longitudinal components of massive gauge bosons are created. In other words, the Goldestone bosons are eaten by the massive gauge bosons. Hence, one can show that in high energy limit, the amplitude for emission or absorption of a longitudinally polarized gauge boson becomes equal to the amplitude for emission or absorption of the Goldestone boson that had been eaten. This statment, well known as the equivalence theorem, is a consequence of a spontaneously symmetry breaking via a gauge invariat manner \cite{eq1,eq2}.

 In order to study the GBET, as an example, we consider the process $W^+_LW^-_L\rightarrow W^+_LW^-_L$  and obtain its amplitude in high energy limit\footnote{Since the masses of new gauge bosons, $M_{W^0}$ and $M_{G^0}$, appear in relevant amplitudes, we need to consider the energies larger than these mass scales which determine the scales of new symmetry breaking mechanisms.}. In addition, we obtain the amplitude of the process $\omega^+\omega^-\rightarrow \omega^+\omega^-$ in which $\omega^\pm$ are the Goldestone bosons eaten by $W^\pm$ during the Higgs mechanism in high energy limit. According to the GBET, they must be equal.


The
paper is organized as follows: In Section 2, we give a brief review of the nSM. In section 3, we consider the $W^+_LW^-_L\rightarrow W^+_LW^-_L$ scattering and give the corresponding amplitude order by order with respect to $s/M_W^2$. In section 4, we give the Lagrangian of Higgs sector after symmetry breaking  and discuss the gauge fixing approach. In addition, the equivalence theorem is mentioned and we write the amplitude of  $\omega^+_L\omega^-_L\rightarrow \omega^+_L\omega^-_L$ and the equivalence theorem is also discussed. Finally, we summarize our results in the last section.
\section{ A brief review of nSM}
According to the NC no-go theorem, the NC gauge group is restricted
to $U_\star(n)$ and permissible  representations are fundamental, anti-fundamental and adjoin representation. In particular, in the case of $U_\star(1)$ only charges $\pm q$ and 0 can be accommodated. Moreover, if we have a gauge group which is a direct product of multi simple groups, the matter fields can be charged under at most  two of them. Therefore, to construct the SM consistent with these restrictions, we have to start with $U_\star(3)\times U_\star(2)\times U_\star(1)$ gauge group which is larger than the usual SM one (as we said this model is called nNCSM). The charge quantization problem can be solved by reduction of the addition symmetry through appropriate Higgs mechanisms \cite{nNCSM}. Fortunately, the number of possible particles which can be accommodated in nNCSM is equal to the number of matter fields in the SM with different hyper-charges. However, it has been shown that the tree level unitarity is violated due to the noncommutative space-time \cite{unitary violation}. Moreover in this paper we are going to study the electroweak spontaneously symmetry breaking at the commutative level. Therefore, in this section we give a brief review of the commutative version of the $U(3)\times U(2)\times U(1)$ model which we called nSM. 

  There are six possible matter fields in this model; left-handed lepton doublet, right-handed charged lepton, left-handed quark doublet, right-handed up quark, right-handed down quark and Higgs doublet, which transform under the SM gauge group, respectively, as follows:
\bea\label{Ll}
{\Psi}^l_L(x)\equiv\Big(\begin{array}{c}
                   {\nu}(x) \\
                   {e}(x)
                 \end{array}
\Big)_L\rightarrow {v}^{-1}(x){V}(x){\Psi}^l_L(x)
\eea
\be\label{Rl}
{e}_R(x)\rightarrow {v}^{-1}(x){e}_R(x)
\ee
\bea\label{Lq}
{\Psi}^q_L(x)\equiv\Big(\begin{array}{c}
                   {u}(x) \\
                   {d}(x)
                 \end{array}
\Big)_L\rightarrow {V}(x){\Psi}^q_L(x) {U}^{-1}(x)
\eea
\be\label{Ru}
{u}_R(x)\rightarrow {v}(x) {u}_R(x) {U}^{-1}(x)
\ee
\be\label{Rd}
{d}_R(x)\rightarrow {d}_R(x) {U}^{-1}(x)
\ee
\bea\label{higgs}
{H}(x)\equiv\Big(\begin{array}{c}
                   {H}^+(x) \\
                   {H}^0(x)
                 \end{array}
\Big)\rightarrow {V}(x) {H}(x),
\eea
where ${v}(x)$, ${V}(x)$ and ${U}(x)$ are, respectively, $U(1)$, $U(2)$ and $U(3)$ gauge transformation.
These transformations along with the following transformation for the gauge fields:
\bea \label{gauget}
{B}_\mu\rightarrow {B}_\mu +\frac i g_1 {v}\p_\mu {v}^{-1}, \\
{W}_\mu\rightarrow {U} {W}_\mu {U}^{-1} +\frac i g_2 {U}\p_\mu {U}^{-1}, \\
{G}_\mu\rightarrow {V} {G}_\mu {V}^{-1} +\frac i g_3 {V}\p_\mu {V}^{-1},
\eea
where ${B}_\mu$, ${W}_\mu$ and ${G}_\mu$ are $U(1)$, $U(2)$ and $U(3)$ gauge fields, respectively, define the $U(3)\times U(2)\times U(1)$ gauge theory including gauge and Yukawa interactions. Of course, conservation of the gauge symmetry in the Yukawa interactions for the up quark leads to the following gauge transformation for the charge conjugated of doublet Higgs field:
\be\label{higgsc}
{H}^C\rightarrow {v}^{-1}(x){V}(x) {H}^C.
\ee

Now, we have to reduce $U(3)\times U(2)\times U(1)$ gauge group to the $SU(3)\times SU(2)\times U(1)$ via appropriate Higgs mechanisms.  In fact, $U(n)$ group can be
decomposed as follows \be U(n)=U_n(1)\times SU(n), \ee where
$U_n(1)$ and $SU(n)$ are the Abelian  and non-Abelian elements, respectively.  In other
words, the elements of $ U(n)$ can be uniquely written as \be
U(x)=e^{i\epsilon_0(x)1_n} e^{i\epsilon_a(x)T^a}. \ee
We can reduce the three $U(1)$'s in $U(3)\times U(2)\times U(1)$ gauge group to one, using two  scalar fields (called
Higgsac).
One of them is used to reduce factors $U_3(1)$ and $U_2(1)$ to an Ablian $U_{32}(1)$ and the other to reduce $U_{32}(1)$ and $U_1(1)$ to the final Ablian $U(1)$.
Consequently, the former has the charges of $U_3(1)$ and $U_2(1)$ and the later has the charges of remained $U(1)$ from the recent symmetry reduction $U_{23}(1)$ and $U_1(1)$.
Hence, the gauge
transformation of the Higgsac field for the first symmetry breaking
is \be\phi_1(x)\rightarrow
 U_{3}(x)\phi_1(x)V_2^{-1}(x),\label{8}\ee
 and for the second one is
 \be\phi_2(x)\rightarrow
 U_{32}(x)\phi_2(x)v_1^{-1}(x),\label{9}\ee
where $U_{3}(x)\in U_3(1)$, $V_2(x)\in U_{2}(1)$, $U_{32}(x)\in U_{32}(1)$ and $v_1(x)\in U_1(1)$.

Until now, we have given the gauge transformations of various fields in the nSM.  The interactions of the gauge, fermionic, Higgs, and Yukawa sectors can be written as follows:
	
{\bf The gauge sector:}
In the nSM, we have gauge field ${B}_\mu(x)$, that is valued in the $u(1)$ algebra, the $u(2)$-valued gauge fields
\be
{W}_\mu(x)=\sum_{A=0}^3{W}^A_\mu(x)\sigma^A,
\ee
where $\sigma^i$, {i=1,2,3}, are $2\times2$ Pauli matrices and $\sigma^0={\bf 1}_{2\times2}$ and $u(3)$-valued gauge fields
\be
{G}_\mu(x)=\sum_{A=0}^8{W}^A_\mu(x)T^A,
\ee
where $T^i$, {i=1,2...8}, are $3\times3$ Gell-Mann matrices and $T^0={\bf 1}_{3\times3}$. Using the gauge field transformations, (\ref{gauget}), one can show that the field strengths defined by
\be
{B}_{\mu\nu}=\p_{[\mu} {B}_{\nu]},
\ee
\be
{W}_{\mu\nu}=\p_{[\mu} {W}_{\nu]}+ig_2[{W}_\mu,{W}_\nu],
\ee
\be
{G}_{\mu\nu}=\p_{[\mu} {G}_{\nu]}+ig_1[{G}_\mu,{G}_\nu],
\ee
transform as
$
{B}_{\mu\nu}\rightarrow {B}_{\mu\nu}
$,
$
{W}_{\mu\nu}\rightarrow V{W}_{\mu\nu} V^{-1}
$ and
$
{G}_{\mu\nu}\rightarrow U{G}_{\mu\nu} U^{-1}
$, respectively. Consequently, the gauge invariant action of gauge sector is:
\be
S_{gauge}=-\frac 1 4 \int d^4x({B}^{\mu\nu}{B}_{\mu\nu}+tr({W}^{\mu\nu}{W}_{\mu\nu})+tr({G}^{\mu\nu}{G}_{\mu\nu})).
\ee
 Now, the gauge group has to be reduced down to $SU(3)\times SU(2)\times U(1)$ which is the SM gauge group. For this purpose, three commutative $U(1)$ subgroups in the $U(3)\times U(2)\times U(1)$ are reduced into one. We chose one of the Higgsacs denoted by $\phi_1$ to be charged under $U(1)$ subgroups of $U(3)$ and $U(2)$ according to (\ref{8}). Therefore, we write the corresponding Lagrangian as follows:
\be
(D_\mu\phi_1)^\dagger(D^\mu\phi_1)+m_1^2\phi_1^\dagger\phi_1-\frac{\lambda_1}{4!}(\phi_1^\dagger\phi_1)^2,
\ee
with
\be
D_\mu\phi_1=\p_\mu\phi_1+\frac i 2 3g_3G^0_\mu\phi_1-\frac i 2 2g_2\phi_1W^0_\mu.
\ee
The remaining $U(1)$ from the recent symmetry reduction along with the initial $U(1)$ are reduced  down to the final $U(1)$. This mechanism is performed through the coupling of the other Higgsac denoted by $\phi_2$ with residual massless gauge field from the previous symmetry reduction, $B^\prime_\mu$,  and the gauge field of the initial $U(1)$, $B_\mu$, given by
 \be
(D_\mu\phi_2)^\dagger(D^\mu\phi_2)+m_1^2\phi_2^\dagger\phi_2-\frac{\lambda_2}{4!}(\phi_2^\dagger\phi_2)^2,
\ee
with
\be
D_\mu\phi_2=\p_\mu\phi_2+\frac i 2 g_0B^\prime_\mu\phi_2-\frac i 2 g_1B_\mu\phi_2,
\ee
where $g_0=2g_23g_3/\sqrt{(2g_2)^2+(3g_3)^2}$. The initial gauge fields $G^0_\mu$, $W^0_\mu$ and $B_\mu$ are written with respect to mass eigenstate fields ${G^0_\mu}^\prime$ and
${W^0_\mu}^\prime$ and hyper-photon field $Y_\mu$ as follows:
\bea\label{vmass1}
\left(
  \begin{array}{c}
    G^0_\mu \\
    W^0_\mu \\
    B_\mu \\
  \end{array}
\right)=R_{23}R_{11^\prime}\left(
                             \begin{array}{c}
                               {G^0_\mu}^\prime \\
                               {W^0_\mu}^\prime \\
                               Y_\mu \\
                             \end{array}\label{111}
                           \right),
\eea
where
\bea
R_{23}=\left(
         \begin{array}{ccc}
           \cos\delta_{23} & \sin\delta_{23} & 0 \\
           -\sin\delta_{23} & \cos\delta_{23} & 0 \\
           0 & 0 & 1 \\
         \end{array}
       \right)\,\,\,\,,\hspace{5mm}\! R_{11^\prime}=\left(
                                     \begin{array}{ccc}
                                       1 & 0 & 0 \\
                                       0 & \cos\delta_{11^\prime} & \sin\delta_{11^\prime} \\
                                       0 & -\sin\delta_{11^\prime} & \cos\delta_{11^\prime} \\
                                     \end{array}
                                   \right),
\eea
and $\delta_{23}$ and $\delta_{11^\prime}$ are defined, respectively, as follows:
\be
\tan\delta_{23}=\frac{2g_2}{3g_3}\,\,\,,\hspace{1cm}\cot\delta_{11^\prime}=\frac{2g_23g_3}{g_1\sqrt{(2g_2)^2+(3g_3)^2}}.
\ee

{\bf Fermionic sector:}
 After symmetry reductions, the gauge fields corresponding to $U(1)$ factors have to be written with respect to the mass eigenstate gauge fields; ${G^0_\mu}^\prime$, ${W^0_\mu}^\prime$, and hyper-photon $Y_\mu$ (see Eq. (\ref{111})). Comparing the hyper-photon coupling with its correspondence in the usual SM, we see that the charge quantization problem inheriting from NC no-go theorem is solved.  We list the Lagrangian of each fermion family and find the relations between the couplings $g_1$ and $g_2$ and the SM hyper-photon coupling $g^\prime$.

{\it 1-Right-handed electron}
\begin{equation}
\bar{{e}}_R\gamma{D}^1_\mu {e}_R=\bar{{e}}_R\gamma^\mu\p_\mu{e}_R-\frac i 2 g_1 \bar{{e}}_R\gamma^\mu{e}_R{B}_\mu.
\end{equation}
 Writing $B_\mu$ in terms of $Y_\mu$ and ${W^0_\mu}^\prime$ from (\ref{vmass1}) and comparing it with the usual SM, we find
\be
-\frac 1 2 g_1\cos\delta_{11^\prime}=g^\prime\frac{Y_{e_R}}{2}=-g^\prime,
\ee
where the hyper-charge of right-handed electron, $Y_{e_R}$, is taken -2.

{\it 2-Left-handed leptons}
\begin{equation}
\bar{{\Psi}}^l_L\ga^\mu {D}_\mu^{1+2}{{\Psi}}^l_L=\bar{{\Psi}}^l_L\ga^\mu\p_\mu{{\Psi}}^l_L+\frac i 2 g_2\bar{{\Psi}}^l_L\ga^\mu{W}_\mu{{\Psi}}^l_L-\frac i 2 g_1\bar{{\Psi}}^l_L\ga^\mu{{\Psi}}^l_L{B}_\mu.\end{equation}

Here, inserting $B_\mu$ and $W^0_\mu$ from (\ref{vmass1}) in terms of $Y_\mu$, ${W^0_\mu}^\prime$ and ${G^0_\mu}^\prime$ and comparing it with the usual SM we can write
\be
\frac{g_2}{2}\cos\delta_{23}\sin\delta_{11^\prime}-\frac{g_1}{2}\cos\delta_{11^\prime}=g^\prime\frac{Y^l_L}{2}=-\frac{g^\prime}{2},
\ee
where $Y^l_L=-1$ is the hyper-charge of left-handed leptons.

{\it 3-Right-handed up quark}
\begin{equation}
\bar{{u}}_R\ga^\mu {D}_\mu^{1+3}{{u}}_R=\bar{{u}}_R\ga^\mu\p_\mu{{u}}_R+\frac i 2 g_1\bar{{u}}_R\ga^\mu{B}_\mu{{u}}_R-\frac i 2 g_3\bar{{u}}_R\ga^\mu T^A{{u}}_R{G}_\mu^A.\end{equation}
Similarly, $B_\mu$ and $G^0_\mu$ in terms of $Y_\mu$, ${W^0_\mu}^\prime$ and ${G^0_\mu}^\prime$ are read from (\ref{vmass1}). Comparing with the SM, one finds
\be
-\frac{g_3}{2}\sin\delta_{23}\sin\delta_{11^\prime}+\frac{g_1}{2}\cos\delta_{11^\prime}=Y_{u_R}g^\prime=\frac 4 3 g^\prime,
\ee
where $Y_{u_R}=\frac{4}{3}$ is the hyper-charge of the up quark.

{\it 4-Right-handed down quark}
\begin{equation}
\bar{{d}}_R\ga^\mu {D}_\mu^{3}{{d}}_R=\bar{{d}}_R\ga^\mu\p_\mu{{d}}_R-\frac i 2 g_3\bar{{d}}_R\ga^\mu T^A{{d}}_R{G}_\mu^A.\end{equation}
The hyper-charge of right-handed quark, $Y_{d_R}=-\frac 2 3$, is obtained as follows:
\be
-\frac{g_3}{2}\sin\delta_{23}\sin\delta_{11^\prime}=Y_{d_R} g^\prime=-\frac 2 3 g^\prime.
\ee
{\it 5-Left-handed quarks}
\begin{equation}
\bar{{\Psi}}^q_L\ga^\mu {D}_\mu^{2+3}{{\Psi}}^q_L=\bar{{\Psi}}^q_L\ga^\mu\p_\mu{{\Psi}}^q_L+\frac i 2 g_2\bar{{\Psi}}^q_L\ga^\mu{W}_\mu{{\Psi}}^q_L-\frac i 2 g_3\bar{{\Psi}}^q_L\ga^\mu{{\Psi}}^q_L{G}_\mu.
\end{equation}
The hyper-charge of the doublet of left-handed quarks, $Y^q_L=\frac 1 3$, is obtained as follows:
\be
\frac{g_2}{2}\cos\delta_{23}\sin\delta_{11^\prime}-\frac{g_3}{2}\sin\delta_{23}\sin\delta_{11^\prime}=Y^q_L g^\prime=\frac 1 3 g^\prime.
\ee

{\bf Higgs doublet:}
 The massive particles in the SM achieve their masses from the interaction with a Higgs doublet,
\be
({D}_\mu{H})^\dagger({D}^\mu{H})-V({H})=(\p_\mu{H}
+\frac{i}{2}g_2{W}_\mu{H})^\dagger(\p^\mu{H}
+\frac{i}{2}g_2{W}^\mu{H})-V({H}),\label{Hcoupling}
\ee
with
\be
V({H})=\mu^2{H}^\dagger{H}+\lambda({H}^\dagger{H})^2,
\ee
From the coupling $Y_\mu$ to $H$ before the spontaneous symmetry breaking one can conclude that
\be
g_2\cos\delta_{23}\sin{\delta_{11^\prime}}=g^\prime.
\ee
Moreover, the Higgs doublet before the symmetry breaking is as follows:
\bea\label{doublet}
H
=\frac{1}{\sqrt{2}}\left(
\begin{array}{c}
	-i(h^1-ih^2) \\
	v+(h+ih^3) \\
\end{array}
\right)
=\frac{1}{\sqrt{2}}\left(
\begin{array}{c}
	\chi \\
	v+(h+i\chi^0) \\
\end{array}
\right),
\eea
where $v=\sqrt{\frac{-\mu^2}{\lambda}}$ is the vacuum expectation value of Higgs $h$. After symmetry breaking the Higgs mass is $M_h=\sqrt{-2\mu^2}$. 

{\bf Yukawa interactions:}
The generation of fermion masses after SM symmetry breaking is due to the gauge invariant Yukawa interactions between SM massive fermions and doublet Higgs
\be
{\bar{\Psi}}_L{\Phi}{\psi}_R+{\bar{\Psi}}_L{\Phi}^C{\psi}_R+c.c.,
\ee
where ${\bar{\Psi}}_L$ and ${\psi}_R$ denote a left-handed doublet fermion and a right-handed fermion in general, respectively. Charged leptons and down quarks achieve their masses through the first term and up quarks do so through the other.
\section{$W^\pm_L$ process}
The SM does not suffer from the tree level unitarity violation. It is as a result of the appropriate Higgs mechanism. Namely, in the case of $W^+_LW^-_L\rightarrow W^+_LW^-_L$ process, the leading order (${\cal O}(s^2/M_W^4)$) is eliminated because the coupling of photon and $Z^0$ to charged gauge bosons $W^\pm$ are $gs_W$ and $gc_W$, respectively. Moreover, the relation $M_{Z^0}c_W=M_W$ along with the above conditions lead to elimination of the sub-leading order (${\cal O}(s/M_W^2)$). In the nSM, there are two new neutral gauge bosons $G^0$  and $W^0$ in addition to the photon and $Z^0$. However, since these new gauge bosons do not couple to $W^\pm$ in commutative space-time, the conditions are similar to the usual SM. Otherwise, in the NC space-time, new neutral gauge bosons contribute, for instance, in $W^+_LW^-_L\rightarrow W^+_LW^-_L$ process as mediators.
It has been shown that while the leading order is removed similar to usual SM, the sub-leading terms remain due to the NC induced interactions and the unitarity is violated \cite{unitary violation}.

As we said, in the commutative level, new gauge bosons $G^0$ and $W^0$ do not couple to $W^\pm$. Therefore, the amplitude of  $W^\pm_L$ in the nSM is equal to one in the SM. Here, we review it order by order.
 The relevant Feynman rules for $W^\pm_L$ process is as follows\footnote{Hereafter, we use the following abbreviations: $g_2\rightarrow g$, $\cos\theta_W\rightarrow c_W$, $\sin\theta_W\rightarrow s_W$, $\sin\theta_{23}\rightarrow s_{23}$, $\cos\theta_{11^\prime}\rightarrow c_{11}$, $\sin\theta_{11^\prime}\rightarrow s_{11}$}:
 \vspace{1.4cm}
 
\bea\label{W4}
\raisebox{-1.1cm}{\includegraphics[scale=0.32]{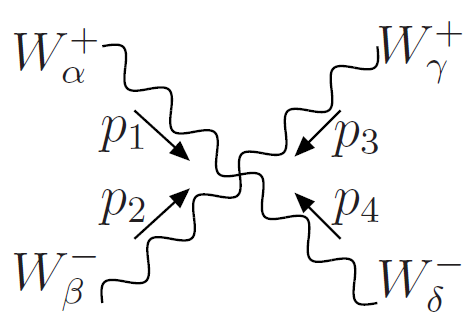}}
\hspace{0mm}=ig^2[2g^{\alpha\gamma}g^{\beta\delta}-g^{\alpha\delta}g^{\beta\gamma}-g^{\alpha\beta}g^{\gamma\delta}],
\eea
\bea
\raisebox{-1.1cm}{\includegraphics[scale=0.32]{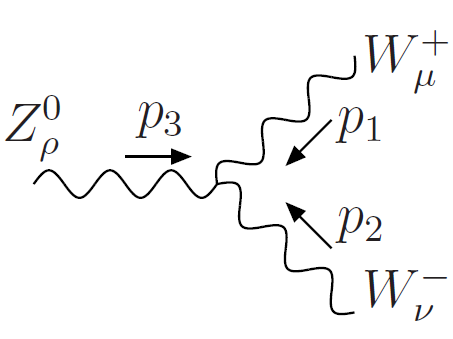}}
\hspace{3mm}=-igc_W
[\hspace{-7mm}&&(p_1-p_2)_\rho g_{\mu\nu}
+(p_2-p_3)_\mu g_{\nu\rho}\\
\hspace{-7mm}&&+(p_3-p_1)_\nu g_{\mu\rho}],\nonumber
\eea
\bea
\raisebox{-1.1cm}{\includegraphics[scale=0.32]{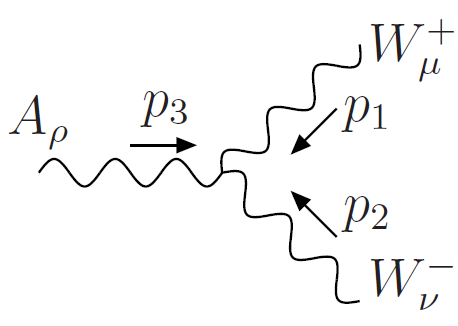}}
=-igs_W[\hspace{-7mm}&&(p_1-p_2)_\rho g_{\mu\nu}
+(p_2-p_3)_\mu g_{\nu\rho}\\
\hspace{-7mm}&&+(p_3-p_1)_\nu g_{\mu\rho}]\nonumber,
\eea
\bea
\hspace{-40mm}
\raisebox{-1.1cm}{\includegraphics[scale=0.32]{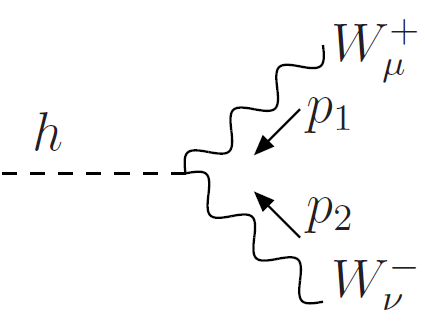}}
\hspace{0mm}=igm_Wg_{\mu\nu}.
\eea

\vspace{15mm}

\begin{figure}
\hspace{0.2cm}
\raisebox{0cm}{\includegraphics[scale=0.45]{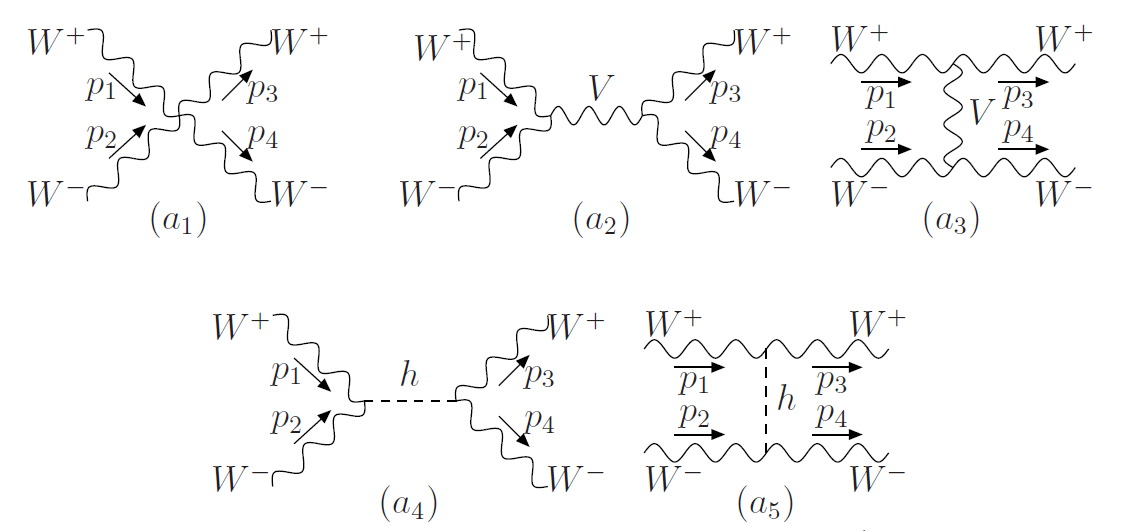}}
\vspace{0mm}
\caption{The leading order Feynman diagrams for $W^\pm_L$ scattering.}
\label{fig1}
\end{figure}

$W_L^\pm$ scattering at the tree level proceeds through the diagrams in Fig. \ref{fig1}. The exchanged vector boson, $V$, in $a_2$ and $a_3$ is either photon or $Z_0$. The leading order ${\cal O}(s^2/M_W^4)$ and sub-leading order ${\cal O}(s/M_W^2)$ vanish consistently with tree level unitarity theorem.

In the high energy limit, the sub-sub leading order, independent of $(s/M_W^2)$, is obtained as follows:
	    	\be
	    	-\frac{ig^2M_h^2}{2M_W^2}+\frac{ig^2}{4c_W^2}\frac{3+\cos^2\theta}{1-\cos\theta}.\label{climit}
	    	\ee
 According to the equivalence theorem, one expects it to be equal to the amplitude of $\omega^+\omega^-\rightarrow\omega^+\omega^-$ in which $\omega^\pm$ are the corresponding Goldestone bosons.
	    	
\section{The Equivalence Theorem and Goldstone Boson Amplitude }
The number of degrees of freedom has to be conserved during the Higgs mechanism. A massless gauge boson has two transverse polarization states. It combines with a scalar Goldstone boson to produce a massive gauge field, which has three polarization states. In other words, the unphysical Goldstone boson is converted to the physical longitudinal polarization of the vector boson. Therefore, one expects that the amplitude for emission or absorption of a longitudinally polarized gauge boson becomes equal to the amplitude for emission or absorption of the corresponding Goldstone boson in high energy limit
\be
{\cal M}(W^+_L,W^-_L,Z_L^0,..)={\cal M}(\omega^+_L,\omega^-_L,\omega^0_L,..)+{\cal O}(M_W^0/s).
\ee
This statement, 
which is known as the equivalence theorem in general, requires the invertibility of the transformations which is the mapping Goldstone bosons to the longitudinal polarization of the vector boson \cite{eq1,eq2}. 

In the nSM, after electroweak symmetry breaking one can write the Lagrangian of the  interactions between the Higgs doublet and the $U(2)$ gauge bosons as follows:
\bea
({D}_\mu{H})^\dagger({D}^\mu{H})=\frac{1}{2}\Big\{|\partial_\mu\chi+(&&\hspace{-6mm}-\frac{ig_2}{2}s_{23}G^0_\mu+\frac{ig_2}{2}c_{23}c_{11}W^0_\mu+\frac{i}{2}(g_2c_W-g^\prime s_W)Z^0_\mu\nonumber\\
&&\hspace{-6mm}+ieA_mu)\chi+\frac{i}{\sqrt{2}}gW_\mu(v+h+i\chi^0)|^2\nonumber
\eea
\vspace{-8mm}
\bea
&&\hspace{+12mm}+|\frac{ig_2}{\sqrt{2}}W^\dagger_\mu\chi+\partial_\mu h+i\partial_\mu\chi^0+(-\frac{ig_2}{2}s_{23}G^0_\mu+\frac{ig_2}{2}c_{23}c_{11}W^0_\mu \nonumber\\
&&\hspace{14mm}-\frac{i}{2}(g_2c_W+g^\prime s_W)Z^0_\mu)(v+h+i\chi^0)|^2\Big\}\label{higgsmechanism}
\eea
In order to remove the unphysical degrees of freedom, according to the Faddeev-Popov approach, we define the following gauge fixing constraints:
\begin{equation}
G_W=\frac{1}{\sqrt{\xi}}(\partial_\mu W^\mu-i\sqrt{2}\xi M_W\chi),\label{gf1}
\end{equation}
\begin{equation}
G_{Z^0}=\frac{1}{\sqrt{\xi}}(\partial_\mu {Z^0}^\mu+\xi M_{Z^0}\chi^0),\label{gf2}
\end{equation}
\begin{equation}
G_{G^0}=\frac{1}{\sqrt{\xi}}(\partial_\mu {G^0}^\mu+\xi s_{23} M_W\chi^0),\label{gf3}
\end{equation}
and
\begin{equation}
G_{W^0}=\frac{1}{\sqrt{\xi}}(\partial_\mu {W^0}^\mu-\xi c_{23}c_{11}M_W\chi^0).\label{gf4}
\end{equation}
It is not possible to satisfy (\ref{gf3}) and (\ref{gf4}), firstly because the gauge freedoms of $G^0$ and $W^0$ are fixed through the Higgac mechanism  and secondly because $\chi^0$ is used to fix the gauge freedom of $Z^0_\mu$. This problem shows that in the nSM, the correspondence between $\chi^0$ and the longitudinal polarization of $Z^0$, $W^0$ and $G^0$ is not invertible and consequently  the GBET is violated.




In the last section, we obtained the amplitude of the $W^\pm_L$ scattering. Now, in order to check the GBET, we obtain the amplitude of the corresponding process in which the external line particles $W^\pm_L$ are replaced by the Goldestone bosons $\omega^\pm$.
The relevant Feynman rules are:
\bea
\hspace{-40mm}
\raisebox{-1.1cm}{\includegraphics[scale=0.32]{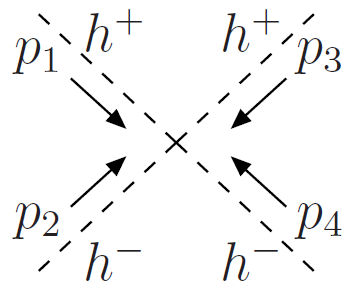}}
\hspace{0mm}=-i\frac{g^2M_h^2}{2M_W^2},
\eea
\be
\hspace{-30mm}
\raisebox{-1.1cm}{\includegraphics[scale=0.32]{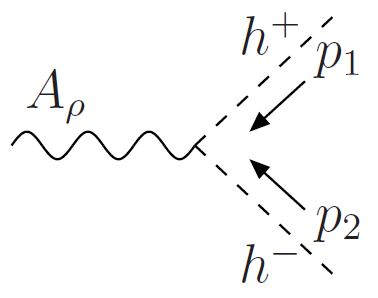}}
\hspace{0mm}=\frac{ig}{2}(s_W+c_{23}s_{11}c_W)(p_2-p_1)_\rho,
\ee
\vspace{5mm}
\be
\hspace{-30mm}
\raisebox{-1.1cm}{\includegraphics[scale=0.32]{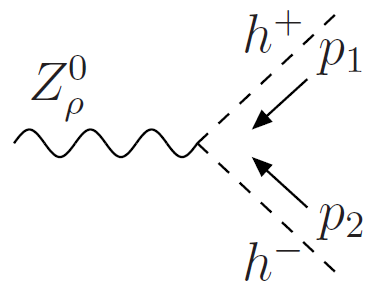}}
\hspace{0mm}=\frac{ig}{2}(c_W-c_{23}s_{11}s_W)(p_2-p_1)_\rho,
\ee
\vspace{5mm}
\be
\hspace{-50mm}
\raisebox{-1.1cm}{\includegraphics[scale=0.32]{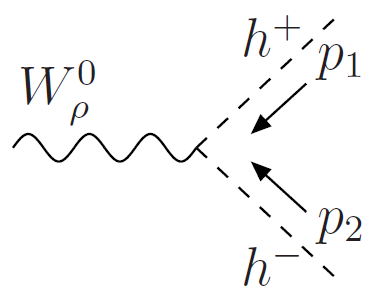}}
\hspace{0mm}=
\frac{ig}{2}c_{23}c_{11}(p_2-p_1)_\rho,
\ee
\vspace{5mm}
\be
\hspace{-50mm}
\raisebox{-1.1cm}{\includegraphics[scale=0.32]{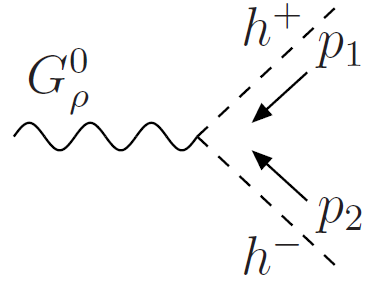}}
\hspace{0mm}=
-\frac{ig}{2}s_{23}(p_2-p_1)_\rho,
\ee
\vspace{5mm}
\be
\hspace{-67mm}
\raisebox{-1.1cm}{\includegraphics[scale=0.32]{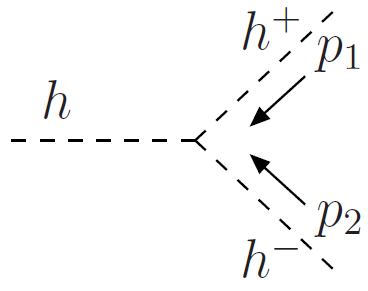}}
\hspace{0mm}=
-i\frac{g^2M_h^2}{4M_W^2} v.
\ee

\vspace{10mm}
We see that $G^0_\mu$ and $W^0_\mu$ interact with $\omega^\pm$ directly. According to the leading order diagrams shown in Fig. \ref{fig2}, we obtain the amplitude of $\omega^+\omega^-\rightarrow \omega^+\omega^-$ in high energy limit as follows:
\begin{figure}
\hspace{0.45cm}
\raisebox{-1.1cm}{\includegraphics[scale=0.45]{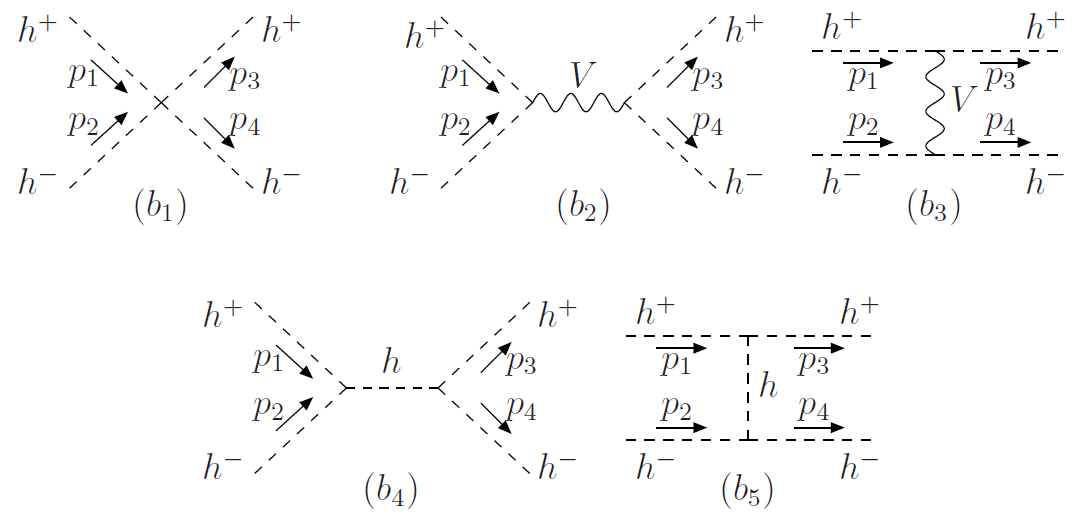}}
\caption{The leading order Feynman diagrams for $\omega^+\omega^-\rightarrow\omega^+\omega^-$}
\label{fig2}
\end{figure}
\be
-\frac{ig^2M_h^2}{2M_W^2}+\frac{ig^2}{4}\frac{3+\cos^2\theta}{1-\cos\theta}.
\label{cgold}
\ee
The coefficient of the second term is different from corresponding expression for the sub-sub-leading order amplitude of $W^\pm_L$ scattering, relation (\ref{climit}). It shows that the GBEQ is violated in the nSM.

\section{Discussion and conclusion}
There is a no-go theorem in NC space-time which restricts the gauge theory \cite{nogo}; gauge groups are limited to $U_\star(n)$, fundamental, anti-fundamental, and adjoint representations are permissible, and  when a gauge group is a direct products of some simple groups $U_\star(n)$, a field can have at most two charges of them. Therefore, transcribing of SM, which is based on $U(1)\times SU(2)\times SU(3)$ gauge group, in NC space-time is problematic. There are two approaches to overcome it \cite{nNCSM,sm}. In one of them which we called nNCSM, one starts with $U_\star(1)\times U_\star(2)\times U_\star(3)$, then using appropriate symmetry reduction, we receive the NC version on SM gauge group \cite{nNCSM}. 
However, 
the tree level unitarity is violated in this model. By studying a various gauge theory with spontaneously symmetry breaking, it was guessed that this violation which is omitted in the commutative limit comes from the Higgsac mechanisms \cite{unitary violation}.

In this paper, we ignored the NC modifications and investigated whether the electroweak symmetry breaking in the nSM is compatible with the Higgsac mechanisms.
In fact, the coupling of massive gauge bosons $G^0$ and $W^0$ to the electroweak Higgs doublet causes the dependence of these mechanisms. We see that the gauge fixing conditions (\ref{gf2}), (\ref{gf3}) and (\ref{gf4}) cannot be compatible simultaneously. This problem leads to the equivalence relation between the Goldestone bosons of Higgs doublet and the longitudinal polarization of massive gauge bosons not to be invertible.  
Hence, the GBET is violated in the nSM. 
For instance, we considered $W^\pm_L$ scattering and obtained its amplitude (Eq. (\ref{climit})). Naturally, it is completely similar to the SM because the new generators commute with the SM ones.  Then we obtained the amplitude for the $\omega^\pm$ scattering (Eq. (\ref{cgold})) ($\omega^\pm$ is the Goldestone boson eaten by $W^\pm$). Comparing Eq. (\ref{climit}) and Eq. (\ref{cgold}), we see that $\omega^\pm$ amplitude is not identical to $W^\pm$ one. Therefore, although we can construct the NCSM base on $U(3)\times U(2)\times U(1)$ such that it is consistent with the NC no-go theorem, the new symmetry reductions are not compatible with the electroweak Higgs mechanism even in the commutative level.

{\bf Acknowledgement:}
The financial support of the University of Qom research council is acknowledged. The author would like to thank R. Moazzemi and Z. Tabatabaie for their reading the manuscript.


\begin{thebibliography}{99}
	
\bibitem{nNCSM} M. Chaichian, P. Presnajder, M. M. Sheikh-Jabbari, and A.
Tureanu, Eur. Phys. J. C{\bf 29}, (2003) 413.

\bibitem{sw}
N. Seiberg and E. Witten, JHEP {\bf 09} (1999)
032.

\bibitem{aash1}
F. Ardalan, H. Arfaei, and M.M. Sheikh-Jabbari, JHEP {\bf 02} (1999) 016.

\bibitem{aash2}
F. Ardalan, H. Arfaei, and M.M. Sheikh-Jabbari, Nucl. Phys. {\bf B576} (2000) 578.


\bibitem{uv/ir}
S. Minwalla, M. van Raamsdonk and N. Seiberg, JHEP {\bf 02}, (2000) 020; A.
Matusis, L. Susskind and N. Toumbas, JHEP {\bf 12}, (2000) 002.

\bibitem{LV}
S.M. Carroll, J.A. Harvey, V.A. Kostelecky, C.D. Lane, and T. Okamoto, Phys.
Rev. Lett. {\bf 87}, (2001) 141601.

\bibitem{CP}
M.M. Sheikh-Jabbari, Phys. Rev. Lett. {\bf 84}, (2000) 5265; P. Aschieri, B. Jurco, P. Schupp, and J. Wess, Nucl. Phys. {\bf B651}, (2003) 45.


\bibitem{nogo}
M. Chaichian, P. Presnajder, M. M. Sheikh-Jabbari, and A. Tureanu,
Phys. Lett. B {\bf 526}, (2002) 132.
\bibitem{sm}
X.~Calmet, B.~Jur\v co, P.~Schupp, J.~Wess and M.~Wohlgenannt, Eur.\
Phys.\ J. {\bf C23}, (2002) 363. 
\bibitem{bypass}
M. Chaichian, P. Presnajder, M.M. Sheikh-Jabbari and A.
Tureanu,Phys. Lett. B {\bf 683},  (2010) 55.


\bibitem{unitary violation}
J. L. Hewett, F. J. Petriello and T. G. Rizzo, Phys. Rev. D {\bf 66}, (2002) 036001.
\bibitem{eq1}
J. M. Cornwall, D. N. Levin, and G. Tiktopoulos, Phys. Rev. {\bf D 10}, (1974) 1145.
\bibitem{eq2}
B. W. Lee, C. Quigg, and H.B. Thacker, Phys. Rev. {\bf D 16}, (1977) 1519.
\end{thebibliography}
\end{document}